\documentclass[epj,twocolumn]{webofc}
\usepackage[varg]{txfonts}   
\usepackage{graphicx}
\usepackage[format=plain,labelformat=default, labelsep=endash, labelfont=bf, margin=0pt, tableposition=top, font={footnotesize}]{caption}
\usepackage[breaklinks=true,colorlinks=true,citecolor=blue,urlcolor=blue]{hyperref} 

\newcommand{\an}{AN,\xspace} 
\newcommand{\aap}{A\&A,\xspace} 
\newcommand{\aj}{AJ,\xspace} 
\newcommand{\apj}{ApJ,\xspace} 
\newcommand{\apjl}{ApJL,\xspace} 
\newcommand{\mnras}{MNRAS,\xspace} 

\newcommand{\pasp}{PASP,\xspace} 

\newcommand{\ie}{i.e.\@\xspace} 
\newcommand{\eg}{e.g.\@\xspace} 

\newcommand{\kp}{\textit{Kepler}\xspace}
\renewcommand{\eqref}[1]{Equation~\ref{#1}}
\newcommand{\fref}[1]{Figure~\ref{#1}}
\newcommand{\sref}[1]{Section~\ref{#1}}

\wocname{epj}
\woctitle{Seismology of the Sun and the Distant Stars 2016}
\begin{document}

\title{\vspace{-3cm}Data preparation for asteroseismology with TESS}
%

\author{\firstname{Mikkel N.} \lastname{Lund}\inst{1,2}\thanks{\email{lundm@bison.ph.bham.ac.uk}} \and
        \firstname{Rasmus} \lastname{Handberg}\inst{2}\thanks{\email{rasmush@phys.au.dk}} \and
        \firstname{Hans} \lastname{Kjeldsen}\inst{2} \and
        \firstname{William J.} \lastname{Chaplin}\inst{1,2} \and \\
        \firstname{J\o rgen} \lastname{Christensen-Dalsgaard}\inst{2}
}

\institute{School of Physics and Astronomy, University of Birmingham, Edgbaston, Birmingham, B15 2TT, United Kingdom
\and
           Stellar Astrophysics Centre, Department of Physics and Astronomy, Aarhus University, Ny Munkegade 120, DK-8000 Aarhus C, Denmark
          }

\abstract{%
  The Transiting Exoplanet Survey Satellite (TESS) is a NASA Astrophysics Explorer mission. Following its scheduled launch in 2017, TESS will focus on detecting exoplanets around the nearest and brightest stars in the sky, for which detailed follow-up observations are possible. TESS will, as the NASA \kp mission, include a asteroseismic program that will be organized within the TESS Asteroseismic Science Consortium (TASC), building on the success of the \kp Asteroseismic Science Consortium (KASC).
Within TASC data for asteroseismic analysis will be prepared by the TASC Working Group 0 (WG-0), who will facilitate data to the community via the TESS Asteroseismic Science Operations Center (TASOC), again building on the success of the corresponding KASOC platform for \kp. Here, we give an overview of the steps being taken within WG-0 to prepare for the upcoming TESS mission.
}
\maketitle

\section{Introduction}
\label{intro}

The Transiting Exoplanet Survey Satellite (TESS) is a NASA Astrophysics Explorer mission \citep{2014SPIE.9143E..20R}, scheduled for launch at the end of 2017 and with a nominal mission duration of 2 years. TESS may be seen as the successor to the NASA \kp mission \citep[][]{2010Sci...327..977B}, and will as \kp search for exoplanets using the transit method --- here, a planet is identified from the dimming produced when it passes in front of its host star. Different from \kp, TESS will focus on the nearest and brightest stars in the sky, allowing for detailed follow-up observations, and will over its nominal mission nearly cover the full sky.
The primary science goal of \kp was to determine the frequency of Earth-like planets in and near the habitable zone of solar-type stars \citep[][]{2010Sci...327..977B}; TESS will instead focus on finding exoplanets smaller than Neptune where a detailed characterization is possible from follow-up observations. 

With the advent of the space-based missions CoRoT \citep[][]{2002ESASP.485...17B} and \kp, the field of asteroseismology has flourished over the last decade \citep[][]{2015EPJWC.10100001G}. The reason for this advancement is that the photometric requirements needed for detecting transiting exoplanets coincide with those needed for asteroseismology, to wit, photometric observations of long duration and high precision. This synergy was realized early on for both the CoRoT and \kp missions, and led for \kp to the formation of the \kp Asteroseismic Investigation (KAI). Via the \kp Asteroseismic Science Consortium \citep[KASC;][]{2010AN....331..966K} this provided direct access to the data from \kp and helped to organize the work within the broad asteroseismic community. 

Building on the success of KASC, the asteroseismic studies in TESS will be organized in the TESS Asteroseismic Science Consortium \citep[TASC;][]{tasc_doc}.
In the following we will focus on the preparation of data from TESS for the sake of asteroseismology.

\section{The TESS mission}
\label{tess}

Over its nominal mission TESS will observe the full sky, starting in the southern hemisphere. The total field of view (FOV) of the four cameras of TESS (each with 4 CCDs) will cover a rectangular slap of the sky spanning $24^{\circ}\times 96^{\circ}$, starting from an ecliptic latitude of ${\sim}6^{\circ}$. A given $24^{\circ}\times 96^{\circ}$ field will be observed for ${\sim}27$-days, corresponding to two orbits of the TESS spacecraft in its highly elliptical 13.7-day Lunar resonances orbit --- we refer to such a field as an observing `Sector'. Given the observing strategy adopted in TESS, some regions will be observed for longer than ${\sim}27$-days. Most notable are the regions within $12^{\circ}$ of the ecliptic poles that will be observed continuously, these are the so-called continuous viewing zones (CVZs).

\begin{figure}
\centering
\includegraphics[width=\columnwidth]{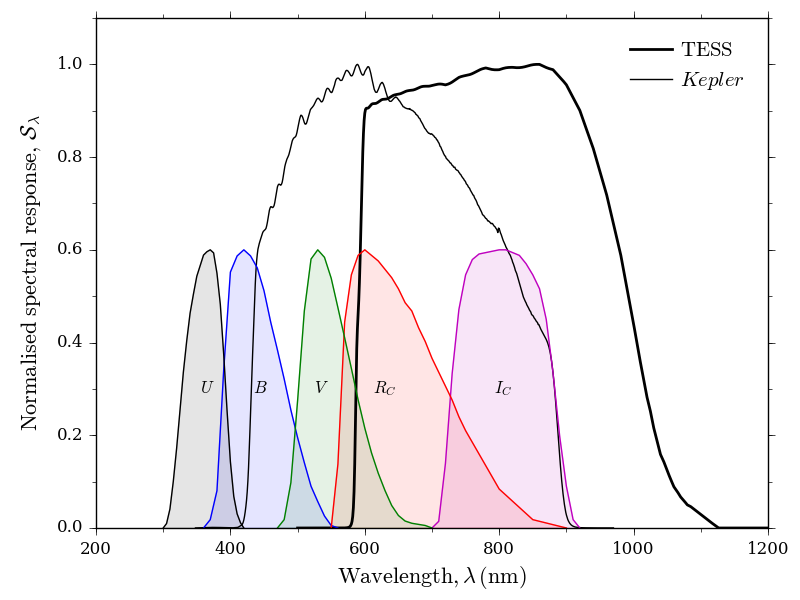}
\caption{Spectral response functions $\mathcal{S}_{\lambda}$ for \kp \citep{VCleve} and TESS \citep{2014SPIE.9143E..20R}, normalised to a maximum of 1. Shown are also the standard Johnson-Cousins $UBVR_CI_C$ photometric systems from \cite[][]{1990PASP..102.1181B}, normalised to maximum values of 0.6.}
\label{fig-0}       
\end{figure}

Observing cadences will come at 20 and 120 seconds, and full-frame-images (FFIs) will be obtained every 30 minutes.
Over the course of the nominal 2 year mission the number of stars observed in 20-sec and 120-sec cadences will exceed $200{,}000$, and data for ${>}20{,}000{,}000$ stars are predicted from the 30-min FFIs. The pixels in TESS are, with a size of $21.1''$, significantly larger than those of \kp, which measured $3.98''$. However, the pixel response function in TESS is very similar to that of \kp, with ${\sim}50\%$ of light contained within 1 pixel, and ${\sim}90\%$ contained within $4\times 4$ pixels. The band-pass of TESS, roughly spanning the interval from $600-1000$ nm and centred on the $I_C$ band, is redder than that of \kp which was centred on the $R_C$ band (see \fref{fig-0}). At short wavelengths the TESS spectral response function is dominated by a long-pass filter transmission, and by the CCD quantum efficiency at long wavelengths. 

Considering the number of stars observed and the larger number of pixels on average devoted to each of these (${\sim}100$ pixels vs. ${\sim}32$ in \kp), the data rate for TESS from 120-sec cadence data will be a factor of ${\sim}13$ that of \kp. If FFIs are included the data rate rises to a factor of ${\sim}25$ that of \kp (Jenkins et al., in prep.).  
Data will be down-linked every $13.7$-days when the TESS spacecraft reaches the perigee of its orbit. Here data will be transferred from TESS to the Deep Space Network (DSN), which will act as the relay for the TESS observations.

\section{The TESS Asteroseismic Investigation}
\label{tai}

As mentioned in \sref{intro}, the \kp Asteroseismic Investigation (KAI) was organized within the broad international community in the KASC.
Building on this, the TESS Asteroseismic Investigation (TAI) will be organized in the TESS Asteroseismic Science Consortium \citep[TASC;][]{tasc_doc}.
Like KASC, the investigations within TASC will be divided between a number of Working Groups (WGs), each of which deals with the utilization of data for a specific
group of objects. Each WG will have two co-chairs who will have the overall responsibility for the running of the WG, and these will be members of the TASC steering committee (SC). 
The TASC-SC, including also the TASC Board, is responsible for the overall running of TASC and will reports to the TESS team on issues pertaining to target selection.
TASC will furthermore organize workshops aiming at target selection, science collaboration and data analysis.

Data and communication platforms for the WGs will be facilitated for TASC via the TESS Asteroseismic Science Operations Center (TASOC)\footnote{\url{tasoc.dk}}, hosted at the Stellar Astrophysics Centre (SAC) at Aarhus University, Denmark. TASOC will furthermore provide long-term storage of all data products.
By and large, TASOC will copy the facilities of the \kp Asteroseismic Science Operations Centre (KASOC)\footnote{\url{kasoc.phys.au.dk}}.
Membership of TASC is open and any member of TASC can apply to become a member of a given WG.
The WG-0 ``\textit{TASOC – Basic photometric algorithms and calibration of time / TASC data products}'' will, as the name suggests, be responsible for maintaining the TASOC portal and the timely provision of data products for the whole of TASC. In \sref{sec-1} below we outline the different main tasks and responsibilities of WG-0.

\section{WG-0 tasks}
\label{sec-1}

WG-0 will have the overall responsibility for delivering analysis-ready data for asteroseismology to TASC in a timely fashion.
For each 27-day pointing, ${\sim}750$ targets at 120-sec cadence, and ${\sim}60$ targets at a 20-sec cadence, will be available for asteroseismology.
WG-0 is, however, committed to the preparation of data for all targets with 120-sec and 20-sec cadences, not only those designated for asteroseismology. 
Additionally, WG-0 will analyse the 30-min FFIs in order to facilitate the detection of oscillations in red giants, SPBs, RR Lyraes, $\beta$ Cep stars, Cepheids, etc., and will also produce light curves for eclipsing binaries. To produce optimally prepared data for the many different types of studies conducted within TASC, WG-0 will maintain close collaborations with the other WGs of TASC. 

The TESS Science Processing Operations Center (SPOC) will process all 120-sec targets in the same manner as done by the Science Operations Center (SOC) for \kp. This includes, for instance, the calibration of pixels, extraction of photometry and astrometry, definition of optimal pixel masks for aperture photometry, correction for systematic errors, etc. --- \ie, an end-to-end analysis.
For FFIs, SPOC is only committed to calibrating and archiving the pixels, while no corrections will be done at all for 20-sec data products (see \sref{sec-1.1}).
Data products from both WG-0 and SPOC will be modelled after those from \kp (Jon Jenkins, private comm.).

\subsection{20-sec-specific data correction}
\label{sec-1.1}

The 20-sec cadence data have been included amongst the cadences employed by TESS primarily for the sake of asteroseimology. The 20-sec cadence will be especially useful for studies of high-frequency oscillators, such as white dwarfs and some main-sequence solar-like oscillators. 
Because this sampling has been introduced for asteroseismology, only fully raw data will be delivered by the TESS team. 
WG-0 will then be responsible for the full calibration and analysis of these data, including basic corrections for 2D black levels; detector gain/linearity; smear; flat-fielding; and the removal of cosmic rays.

\subsubsection{Cosmic rays}
\label{sec-1.1.1}

\begin{figure}
\centering
\includegraphics[trim={4cm 1.7cm 4.5cm 0},clip, width=0.45\columnwidth]{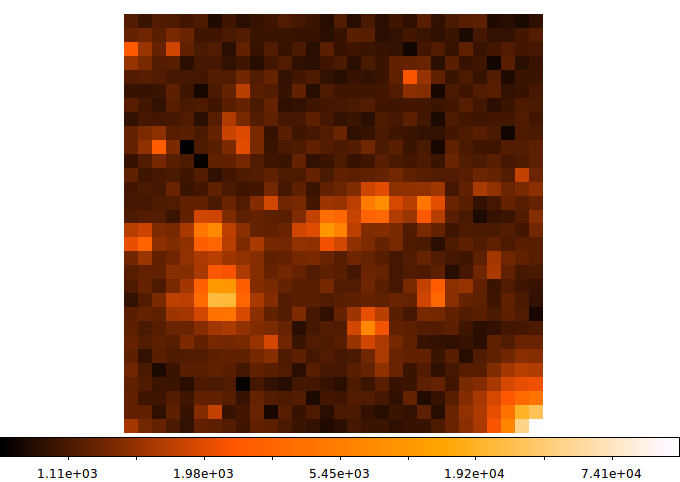}
\hfill
\includegraphics[trim={4cm 1.7cm 4.5cm 0},clip, width=0.45\columnwidth]{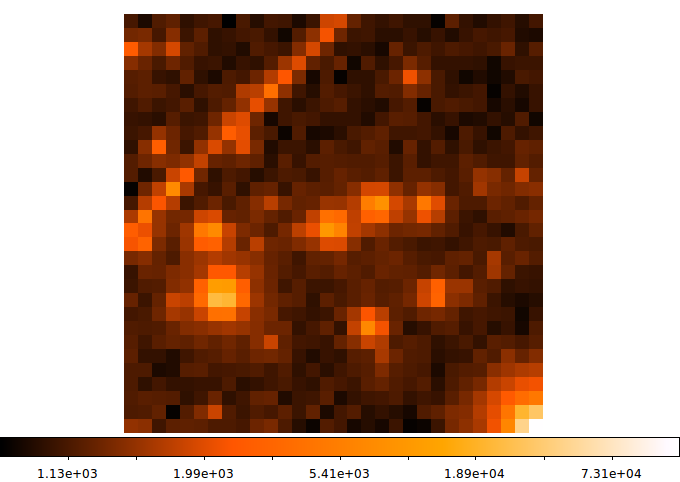}
\caption{Potential effect of CRs in TESS. Shown are two 20-sec cadences of the same simulated pixel-field, one of which (right) are affected by a CR producing a trail impacting many pixels.}
\label{fig-1}       
\end{figure}

For 120-sec data and the 30-min FFIs, cosmic-ray (CR) signals will be mitigated on-board before the cadences are created from the 2-sec integrations in TESS.
The idea for this mitigation is, at the time of writing, to identify outliers in the 2-sec light curves of individual pixels. If a given pixel is found to be affected by CRs, the identified 2-sec samplings are removed before the data are co-added to the 120-sec and 30-min cadences.
Given that the 20-sec data will only consist of 10 such 2-sec integrations, it has been decided that removing the CRs from the co-added data on ground is more optimal. For every 20-sec cadence there is a ${\sim}1.7\%$ chance per pixel for a CR hit.
WG-0 will before launch need to identify suitable methods for such a correction.
It is worth noting that CRs in TESS will impact the photometry in a manner quite different to that in \kp, because of the difference in the pixels between TESS and \kp. In TESS, the pixels have a width of $\rm 15\mu m$ and a depth of $\rm 100\mu m$, whereas \kp use pixels with a width of $\rm 27\mu m$ and a depth of $\rm 15\mu m$. The reason for this choice is the desire for a high spectral response at long wavelengths (\fref{fig-0}), which requires significantly deeper pixels due to the quantum efficiency of the detector material.    
The deeper pixels, however, means that the cross-section of the detector for an incoming CR is much larger than in \kp. \fref{fig-1} shows a simulated pixel field at two different 20-sec cadences, where one (right panel) is affected by a CR. Where such an event in \kp would likely only have affected a single pixel, it can in TESS produce a trail which impacts many pixels.     

\subsection{Sky backgrounds}
\label{sec-1.2}

For 20-sec data and FFIs WG-0 will need to estimate sky-background (SB) levels.
The non-instrumental SB is mainly composed of the contribution from the diffuse background of unresolved stars and galaxies and the sky glow from Zodiacal light, which depends especially on ecliptic latitude \citep[see, \eg,][]{2015ApJ...809...77S}. Before launch, WG-0 will work towards a proper and robust estimation of the SB for the highly diverse fields covered by TESS, going from near-ecliptic to polar and from very sparse to very dense (including regions containing stellar clusters, see \fref{fig-2}).

\subsection{Extracting photometry}
\label{sec-1.3}

WG-0 is committed to extracting light curves for all possible sources in the 20-sec, 120-sec, and 30-min FFI data.
As mentioned in \sref{intro}, this will over the course of the nominal 2-year mission amount to ${>}200{,}000$ star from 20-sec and 120-sec cadences, and ${>}20{,}000{,}000$ stars from the 30-min FFIs.
This number of targets, coupled with the requirement of a timely processing, means that the pipeline constructed for this task will need to be both fast and robust. 
The pipeline will also have to be flexible in terms of its ability to process very diverse fields, including dense fields close to the ecliptic, nebulous regions with high contamination from the SB, and open as well as globular clusters (\fref{fig-2}). It will be especially interesting in the pre-flight tests (\sref{sec-2}) to see what can be expected for studies of star clusters given the relatively large TESS pixels.

Many methods exist for extracting photometry from CCD images, including aperture, point-spread-function (PSF), and so-called optimal photometry \citep[][]{1987PASP...99..191S,1989PASP..101..616H,1998MNRAS.296..339N,2005PASP..117.1113M}. Some of these have already been adapted, or extended upon, for the \kp and K2 missions \citep[][]{2010ApJ...713L..97B,2015ApJ...806...30L,2015MNRAS.447.2880A,2016MNRAS.456.1137L,2016MNRAS.tmp.1279N}. 
Each of the methods have their pros and cons --- aperture photometry is by far the simplest and fastest method, but deciding the optimal size and shape of the aperture is not always straightforward, and it is far from optimal for dense and crowded regions; optimal photometry can provide a more accurate extraction, but it is slower, requires knowledge (albeit not particularly accurate) of the PSF, and is still not optimal for dense and crowded regions; PSF photometry is optimal for dense and crowded regions, but requires accurate knowledge of the PSF and is again slower than aperture photometry. Concerning the PSF, it is worth noting that the TESS PSF will include both off-axis aberrations and chromatic aberrations arising both from the refractive elements of the TESS camera and from the deep-depletion CCDs, absorbing redder photons deeper in the silicon.

All these aspects of the different possible methods must be considered in a final pipeline --- ideally, each method should be thoroughly tested on realistic simulated data, considering here also the hardware requirements that will be needed to keep up with the high data rates of TESS.   
In the end, light curves may well have to be extracted with a range of different methods, depending on the type or crowding of the field under study. Another option might be to run several methods for all fields, with the optimum choice of extracted photometry being made only after the fact.

\subsection{Light curve preparation}
\label{sec-1.4}

Following the extraction of raw light curves from pixel data, WG-0 will for each star produce an analysis-ready light curve for asteroseismology, corrected for any instrumental features.
From \kp we know that instrumental features can come in many forms \citep[][]{2010ApJ...713L.120J,2011MNRAS.414L...6G,2014MNRAS.445.2698H}, including jumps from drops in pixel sensitivity, or from differences in sensitivities between the CCDs that a given star might land on. Such shifts in CCD position happened every Quarter in \kp, and will also occur with TESS for stars with observing durations exceeding the ${\sim}27$ days of an observing Sector; secular changes from variations in focus (\eg from a change in solar heating of the spacecraft), or drifts either in pointing or from differential velocity aberrations; abrupt changes after safe-mode events or data down-links (which will happen every 13.7-days with TESS); transient events such as the Argabrightening events found in \kp \citep[][]{2011SPIE.8151E..17W}, CRs, or from momentum dumps in the reaction wheels orienting the spacecraft.

Currently, we can only speculate about the instrumental features that will be found in TESS, but it is near certain that some features will be found.
The instrumental features that might be found cannot simply be rectified in the same manner for all types of stars under study by TASC (including solar-like oscillators, RR Lyraes, white dwarfs, eclipsing binaries, etc.).
When observing a given star, the observed signal will be a mix of physical and instrumental contributions. Given that the time scales, amplitudes, and phase stability of the physical component will depend on the type of star observed, and thus also on its overlap with the instrumental signals, the method for isolating the instrumental contribution and preserving the astrophysical signal will in effect also depend on the stellar type.      

The idea in WG-0 is to build on the collective knowledge of the community by bringing together people with expertise on the data preparation for different stellar types \citep[see, \eg,][]{2011MNRAS.414L...6G,2011MNRAS.411..878K,2012PASP..124..963K,2013EPSC....8..599D,2014MNRAS.445.2698H,2016AJ....151...68K}. Many methods for rectifying light curves for analysis were developed during the \kp mission, and more recently for the re-purposed K2 mission \citep[see, \eg,][]{2014PASP..126..398H,2014PASP..126..948V,2015ApJ...806...30L,2016MNRAS.455L..36P,2016van.cleve.pasp,2016MNRAS.459.2408A}.
WG-0 will develop a data-correction pipeline that adopts a star-based approach to the mitigation of instrumental effects; this will build on pipelines developed during the \kp mission for specific types of stars. For the pipeline it is worth keeping the high data rate of TESS in mind --- not only should the pipeline be robust and able to handle a diverse range of stellar types, it should also be fast enough to allow for a timely facilitation of processed data.
Several versions of light curves will be available via TASOC for a given star, including a raw uncorrected light curve; a `standard' light curve where the correction method adopted is the same for all stars; and a star-type customized light curve (based on the inputs and request of the TASC community).

\subsection{Absolute timing}
\label{sec-1.5}

The TESS on-board clock should be accurate and stable to better than ${\sim}5$ msec. To obtain a similar accuracy on the time stamps in Barycenter Julian Days (BJD) in the Earth frame, the correction to the light travel time between the spacecraft and the DSN should be accurate to the same level. This will be achieved from knowing the 3D-position of the TESS spacecraft in space to a high level of accuracy (1500 km, corresponding to a light travel time of 5 msec). 
However, delays may occur in the ground system (\eg after data down-links or safe mode event) that cannot be accounted for without an independent assessment of any temporal shifts.

For the sake of ground-based follow-up observations, \eg of transiting exoplanet hosts, it is naturally worth knowing the absolute time stamps of the data. 
Requirements on the accuracy of the absolute timing comes also from asteroseismology \citep[][]{tasc_time}:
\begin{itemize}
\item[$\circ$] To reach the highest possible photometric quality from 120-sec observations, and the photon noise limit for the brightest stars, the absolute photometry needs to be accurate and stable to better than 5 msec. 
\item[$\circ$] To reach the theoretical accuracy of high-amplitude coherent oscillations one needs the time at which each exposure is obtained to be very accurate over the period of an ($27$-day) observing Sector. For coherent pulsation modes this requires that the length of exposure is accurate over an observing Sector to better than 5 msec.
\item[$\circ$] To allow comparisons between ground-based observations with those from TESS, one needs to be able to estimate the absolute time of a given photometric data point and establish a stable reference (e.g. central time of a given observation). For coherent pulsation modes the absolute time (in HJD/BJD) should be known to better than 0.5 sec; for solar-like oscillations the required accuracy is better than 1 sec over a ${\sim}10$ day period.
\end{itemize}
For the calculations leading to these estimates see \cite[][]{tasc_time}.

The TESS team will make the corrections based on calculated light travel times; WG-0 is then committed to making independent checks of the absolute time stamps. 
The regular calibrations will be achieved by performing contemporaneous observations between TESS and ground-based facilities of several objects with photometry varying rapidly in time, such as bright, deep, detached eclipsing binaries.
The absolute time shift, if any, can then be determined by cross-correlating the contemporaneous time series. The ideal objects for these checks will be found in the CVZs of TESS.

The work on the absolute timing issue will be handled by a dedicated sub-group of WG-0.
As the checks of absolute times should be done regularly, and possibly after any data down-link or safe-mode event, the sub-group will have to be able to respond and obtain ground-based data on short notice.
WG-0 will here depend on members of the TASC community with access to ground-based facilities.

\subsection{Stellar classification}
\label{sec-1.6}

An additional sub-group will be formed under WG-0 to perform stellar classification of stars observed with TESS.
The classification is important to select the proper course of action in rectifying a given light curve for asteroseismic studies (\sref{sec-1.4}).
WG-0 will conduct studies of the best classification of stars from the raw photometric data from TESS --- this will be achieved using techniques from machine learning \citep[see, \eg,][]{2009A&A...506..519D,2016MNRAS.456.2260A,2016MNRAS.459.3721B,2016MNRAS.457.3119D}, which will be tested on simulated TESS data before launch. 
\begin{figure*}[!tp]
\centering
\includegraphics[width=0.44\textwidth]{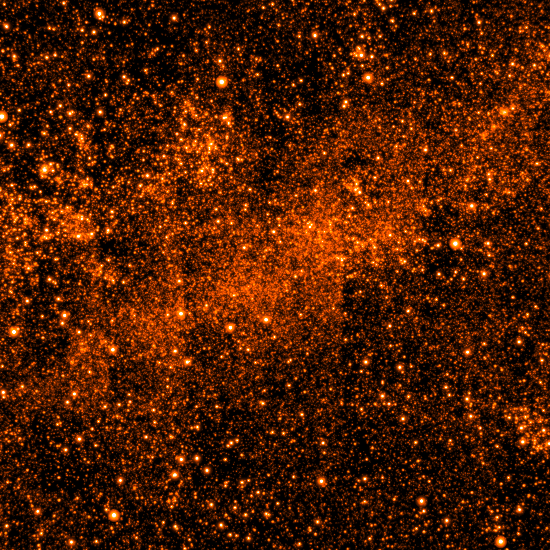}
\hfill
\includegraphics[width=0.44\textwidth]{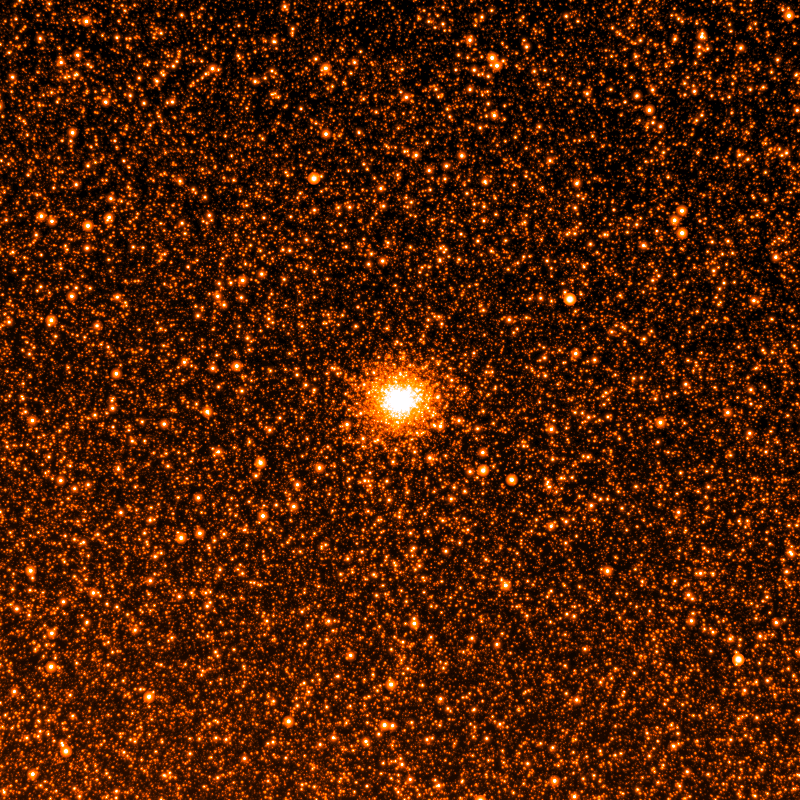}
\caption{Simulated pixels fields from \texttt{SPyFFI} of the Large Magellanic Cloud (LMC; left) and the globular cluster $\omega$ Centauri (NGC 5139; right).}
\label{fig-2}       
\end{figure*}

\section{Pre-flight tests}
\label{sec-2}

In order for WG-0 to be able to construct a data processing pipeline that is ready when the first data from TESS are received, numerous tests will be conducted on simulated data (\sref{sec-2.1}).

\subsection{Pixel-data simulation}
\label{sec-2.1}

Pre-flight analysis will be performed on simulated TESS pixel data made using the ``Spiffy Python for Full Frame Images'' (\texttt{SPyFFI}) simulator.
The simulator was created at the Massachusetts Institute of Technology (MIT) by Zachory K. Berta-Thompson (private comm.). 
As the name suggests, \texttt{SPyFFI} is a Python-based code for simulating TESS pixel data, including FFIs. 

To simulate a given field, \texttt{SPyFFI} uses a user-specified input catalogue with stellar positions and magnitudes. The UCAC4 \citep[][]{2013AJ....145...44Z} catalog is currently used, but eventually the TESS Input Catalog \citep[TIC;][]{TIC} will be adopted. \texttt{SPyFFI} includes realistic models for the TESS pixel response, differential velocity aberration, cosmic rays, spacecraft jitter, focus changes, and sky backgrounds (and the parameters of all of these contributions can be adjusted to test methods from best- to worst-case scenarios).
\fref{fig-2} gives examples of two simulated TESS pixel fields, one of the Large Magellanic Cloud (LMC) and one of the $\omega$ Centauri globular cluster.

\texttt{SPyFFI} furthermore has the option of assigning a simulated light curve to a given star in a given field.
These light curves can include transits, eclipses, spot modulations, and/or oscillations. 
The light curves with solar-like oscillations and granulation signals are produced using the asteroFLAG simulator \cite[][]{2008AN....329..549C}; light curves for classical oscillators have been constructed 
with frequencies, phases, and amplitudes from such stars observed by \kp (Vichi Antoci and Steven Kawaler, private comm.). 

\subsection{T'DA workshop series}
\label{sec-2.2}

To address the issues of TESS data preparation for asteroseismology, WG-0 is organizing the workshop series ``TESS Data for Asteroseismology'' (T'DA).
The idea is to bring together people from the broad community, who either have expertise from missions such as \kp or CoRoT, or who are students planning to work on data analysis issues.
The T'DA series is planned to include, at least, workshops dedicated to (1) extracting light curves from pixel data; (2) correcting light curves for the optimal output from asteroseismic analysis; and (3) stellar classification.
The first workshop (T'DA1), entitled ``From Pixels to Light Curves'', will be held at the University of Birmingham, UK, from 31st Oct. to 2nd Nov. 2016.



\subsection*{Acknowledgements}
{\small The authors would like to thank the organisers of the ``Seismology of the Sun and the Distant Stars 2016 --- Using Today’s Successes to Prepare the Future'' (SpaceTK16) conference, a joint TASC2-KASC9 Workshop -- SPACEINN \& HELAS8 Conference, where MNL presented a talk on the contents of these proceedings.
Funding for the Stellar Astrophysics Centre (SAC) is provided by The Danish National Research Foundation (Grant DNRF106). MNL acknowledges the support of The Danish Council for Independent Research | Natural Science (Grant DFF-4181-00415). WJC acknowledges the support of the UK Science and Technology Facilities Council (STFC).
}


\end{document}